\documentstyle[12pt,fleqn]{article}
\def\beeq{\begin{equation}}
\def\eneq{\end{equation}}
\def\beeqa{\begin{eqnarray}}
\def\eneqa{\end{eqnarray}}

\addtocounter{section}{-1}
\begin{document}

\begin{center}

\vspace{2cm}

{\large {\bf {Disorder effects and electronic conductance\\
in metallic carbon nanotubes
} } }

\vspace{1cm}

{\rm Kikuo Harigaya\footnote[1]{E-mail address: 
\verb+harigaya@etl.go.jp+; URL: 
\verb+http://www.etl.go.jp/+\~{}\verb+harigaya/+}}

\vspace{1cm}

{\sl Physical Science Division,
Electrotechnical Laboratory,\\ 
Umezono 1-1-4, Tsukuba 305-8568, 
Japan}\footnote[2]{Corresponding address}\\
{\sl National Institute of Materials and Chemical Research,\\ 
Higashi 1-1, Tsukuba 305-8565, Japan}\\
{\sl Kanazawa Institute of Technology,\\
Ohgigaoka 7-1, Nonoichi 921-8501, Japan}

\end{center}

\vspace{1cm}

\noindent
{\bf Abstract}\\
Disorder effects on the density of states and electronic
conduction in metallic carbon nanotubes are analyzed
by a tight binding model with Gaussian bond disorder.
Metallic armchair and zigzag nanotubes are considered.
We obtain a conductance which becomes smaller by the 
factor $1/2 \sim 1/3$ from that of the clean 
nanotube.  This decrease mainly comes from lattice 
fluctuations of the width which is comparable to thermal
fluctuations.   We also find that suppression of 
electronic conductance around the Fermi energy due to
disorder is smaller than that of the inner valence
(and conduction) band states.  This is a consequence of 
the extended nature of electronic states around the Fermi 
energy between the valence and conduction bands, and is
a property typical of the electronic structures of 
metallic carbon nanotubes.

\vspace{1cm}

\pagebreak

\section{Introduction}

Recently, carbon nanotubes with cylindrical graphite 
structures have been intensively investigated.  Many interesting 
experimental as well as theoretical researches have been 
performed (see reviews [1,2] for example), and the fundamental
metallic and semiconducting behaviors of single wall nanotubes
predicted by theories [3-8] have been clarified in tunneling 
spectroscopy experiments [9,10].

Measurements of transport properties of single and multi wall 
nanotubes depend largely on various factors including 
sample quality, experimental conditions, and so on. 
Such factors make experimental interpretations difficult.
However, several interesting fundamental properties have
been found.   Magnetoconductance depending on magnetic field
has been measured for multi wall nanotubes, and has been 
interpreted in terms of two dimensional weak localization
and universal conductance fluctuations in mesoscopic 
conductors [11].  Single electron tunneling experiments
of ropes of single wall carbon nanotubes have been performed,
and discrete energy levels of nanotubes between metallic 
contacts contribute to the single electron tunneling 
processes [12].  Electron correlation effects in the 
single electron tunneling have also been observed [13].  
Furthermore, quantized conductance by the multiples of 
the unit conductance $2e^2/h = (12.9 {\rm k}\Omega)^{-1}$ 
has been measured for multi wall carbon nanotubes [14].

In view of the experimental developments of conduction properties, 
it is interesting to investigate basic properties of single
and multi wall carbon nanotubes theoretically. For 
conduction properties, interplay between disorder and
possibility of long ballistic conduction has been
studied [15].  The Landauer formula [16] has been 
used to calculate quantum ballistic transport properties 
of nanotubes for example in [17,18].  Quantum tunneling of 
carbon-nanotube-based quantum dots has been studied in [19].

In this paper, we would like to try to apply the Thouless
formula [20,21] differently from the works in literatures
[17,18], in order to look at possible decrease of the 
electronic conductance by a bond disorder potential.
The origin of the bond disorder potential is the 
thermal fluctuations of phonons mainly.  This idea 
has been used in the discussion of disorder effects on 
the polaron excitations in doped C$_{60}$ systems [22].
We use a tight binding model with Gaussian bond disorder,
and finite systems with quite large metallic carbon 
nanotubes are diagonalized numerically in real space.  
The electronic conductance calculated by the Thouless 
formula is averaged over random samples of disorder.  
The strength of bond disorder is changed within the width
whose magnitude is typical to thermal fluctuation of 
phonons as estimated in ${\rm C}_{60}$ and carbon 
nanotubes [8,22].

We will discuss disorder effects on the density of states and 
electronic conduction in metallic carbon nanotubes, i.e.,
armchair and metallic zigzag nanotubes [23].  The conductance 
at the Fermi energy becomes smaller by the factor $1/2 \sim 1/3$ 
from that of the clean nanotube.  This decrease mainly comes 
from lattice fluctuations of phonons.  The suppression of electronic 
conductance around the Fermi energy due to disorder is 
smaller than that of the main part of the valence (and 
conduction) band states.  This is due to the extended nature 
of electronic states around the Fermi energy

This paper is organized as follows.  In Sec. II, the tight
binding model and the numerical calculation method are 
explained. Sections III and IV are devoted to the results of 
metallic armchair and zigzag nanotubes, respectively.  The 
average conductance at the Fermi energy is discussed in 
Sec. V.  The paper is summarized in Sec. VI.

\section{Model}

Figure 1 shows the way of making general carbon nanotubes
and the notations.  The lattice points in the honeycomb 
lattice are labeled by the vector $(m,n) \equiv m {\bf a} 
+ n {\bf b}$, where ${\bf a}$ and ${\bf b}$ are the unit 
vectors.  Any structure of nanotubes can be produced by 
contacting the origin $(0,0)$ with one of the $(m,n)$ 
vectors after rolling up the plane of the honeycomb lattice 
pattern.  This vector is used as a name of each nanotube.  
The electronic structures of a simple tight binding model 
with nearest neighbor hopping interactions have been found 
theoretically.  When the origin of the honeycomb lattice 
pattern is so combined with one of the open circles as to 
make a nanotube, the metallic properties will be expected 
because of the presence of the Fermi surface.  This case 
corresponds to the vectors where $m-n$ is a multiple of three.  
If the origin is combined with the filled circles, there 
remains a large gap of the order of 1 eV.  This type of 
nanotubes is a semiconductor.

In Fig. 1, the most characteristic pattern in the two 
dimensional graphite plane, where the coupling between
bond alternations and electrons is present, is superposed.
We have refered to this pattern as the Kekul\'{e} structure [7,8].  
The short and long bonds are indicated by the thick and 
normal lines, respectively.  This pattern is commensurate 
with the lattice structure for the metallic $(m,n)$ nanotubes,
and this bond alternation pattern is realized in the
adiabatic approximation.  However, the strength of the
bond alternations is of the order smaller than the experimentally
accessible magnitude [8].  The Kekul\'{e} bond alternation 
pattern will be easily distorted by the fluctuations of 
the phonons from the classical values.  Therefore, we can
neglect the bond alternation effects in order to discuss
disorder effects on electronic conductance of metallic
carbon nanotubes.

Our model Hamiltonian is:
\beeq
H = - t \sum_{\langle i,j \rangle, \sigma} 
(c_{i,\sigma}^\dagger c_{j,\sigma} + {\rm h.c.})
+ \sum_{\langle i,j \rangle, \sigma} \delta t_{i,j}
(c_{i,\sigma}^\dagger c_{j,\sigma} + {\rm h.c.}).
\eneq
The first term is the tight binding model with the nearest 
neighbor hopping interaction $t$; the sum is taken over 
neighboring pairs of lattice sites $\langle i,j \rangle$ 
and spin $\sigma$; $c_{j,\sigma}$ is an annihilation operator
of an electron with spin $\sigma$ at the site $i$.  The 
second term is the bond disorder model, and the hopping 
interaction $\delta t_{i,j}$ obeys the Gaussian distribution 
function
\beeq
P(\delta t)=\frac{1}{\sqrt{4\pi}t_s}
{\rm exp}[-\frac{1}{2}(\frac{\delta t}{t_s})^2]
\eneq
with the strength $t_s$.

A finite system with the quite large system size $N$ of 
metallic carbon nanotubes is diagonalized numerically.  
In this paper, we take $N=4000$ for (5,5) and (10,10)
nanotubes, and $N=3600$ for the (9,0) nanotube.  The 
electronic conductance calculated by the Thouless formula 
is averaged over 100 samples of disorder.  In order to 
look at dependences on disorder strengths $t_s$, we have 
not taken a larger number of samples.  However, dependences 
on $t_s$ are fairly smooth.  So, we can discuss typical 
behaviors of disorder effects, even though numerical error 
bars remain with a certain magnitude.  The quantity $t_s$ is 
changed within $0 \leq t_s \leq 0.3t$  Note that quantities
with the dimension of energy $E$ are measured in units of 
$t$ ($\sim 2$ eV) in this paper.  The typical magnitude 
originating from thermal fluctuation of phonons has been 
estimated as about $t_s \sim 0.15t$ for ${\rm C}_{60}$ 
and carbon nanotubes [8,22].  Therefore, the above range 
of variations seems reasonable.

\section{Armchair nanotubes}

Two characteristic structures of the armchair nanotubes with 
(5,5) and (10,10) geometries are investigated.  The number 
of lattice sites in the unit cell is 20 or 40 for each nanotube.
Thus, the system with $N=4000$ has 200 or 100 unit 
cells.  The diameter of the (5,5) nanotube is similar
to that of C$_{60}$, because this nanotube can be made 
from C$_{60}$ by iterative addition of 10 carbons 
between two hemisphere of C$_{60}$ [7].  This diameter
is the smallest one observed in experiments.  The diameter 
of the (10,10) nanotube is larger than that of the 
(5,5) nanotube, but most of the observed nanotubes 
have diameters similar to that of the (10,10) nanotube.

First, we discuss (5,5) armchair nanotubes.  Figure 2 
shows the density of states (DOS) and electronic conductance of
the clean nanotube.  Figures 2 (a) and (b) show the entire
DOS and the enlarged DOS of the energy region $-1.5 t \leq E
\leq 1.5t$, respectively.  Figure 2 (c) shows the electronic 
conductance of the energy region as in Fig. 2 (b).  We find 
several peaks of the conductance at the energies $\pm 0.6t$, 
$\pm 1.0t$, and $\pm 1.3t$ in Fig. 2 (c) at the corresponding 
peaks of the DOS of Fig. 2(b).  This is the usual results by 
the Thouless formula, though the peaks are somewhat less sharp.
The conductance at the energies less than $-0.6t$ and larger 
than $0.6$, i.e. in the inner valence (conduction) band regions,
is around 2.0 in the unit of $2e^2/h$.  Here, $e$ is
the unit charge, $h$ is the Planck constant, and the factor 2
comes from spin degeneracy.  And, the conductance is around 
the value 1.0 ($2e^2/h$) in the energy region $-0.6 t \leq 
E \leq 0.6t$.

Figure 3 shows one example of the DOS and electronic 
conductance of the (5,5) nanotube with the disorder 
strength $t_s = 0.15t$. In Figs. 3 (a) and (b), the 
strong one-dimensional peaks in the DOS are broaden 
and suppressed.  However, the flat DOS near 
the Fermi level does not change so much, because 
the DOS in these energies is nearly the same.  Figure 3 (c) 
shows the conductance in the energy region $-0.6 t \leq
E \leq 0.6t$.  The conductance at the energies less than 
$-0.6t$ and larger than $0.6t$, i.e. in the inner conduction 
(valence) band regions, is around 0.1 in the unit of $2e^2/h$.
This magnitude is one order smaller than that in the clean 
system.  On the other hand, the conductance is around 
the value 0.3 ($2e^2/h$) in the energy region $-0.6 t \leq 
E \leq 0.6t$.  This value is of the same order of magnitude
as that of the clean system.  Therefore, we find that the
conductance in the inner valence (conduction) band regions
is easily suppressed by the disorder.  However, the 
conductance near the Fermi level is not suppressed so much, 
since conduction and valence bands are mutually connected 
in the metallic carbon nanotube, and therefore the Fermi 
level is located just at the center of the whole energy 
bands.  Then, the disorder effects are smallest at the 
center of the entire energy bands, which means the extended 
nature of electronic states around the Fermi energy.

Next, we look at the results of the (10,10) nanotube,
whose diameter is typical for the observed nanotubes.
Figures 4 and 5 are for the clean nanotube and the
nanotube with the disorder $t_s = 0.15t$, respectively.
As the diameter of the nanotube becomes larger, the
number of peaks in the one dimensional DOS
increases in Figs. 4 (a) and 5 (a) from that of
Figs. 2 (a) and 3 (a).  As the DOS of the clean system 
is flat in the clean system of Fig. 4 (b), the DOS of 
the system with disorder does not change apparently as 
shown in Fig. 5 (b).  The conductance of the inner
valence (conduction) band regions is suppressed by 
one order of magnitudes when disorder is taken into
account as shown by Figs. 4 (c) and 5 (c).  However,
the magnitude of the conductance around the Fermi
energy remains with the similar value.  Such a 
qualitative difference gives rise to larger conductance
around the Fermi energy than that of the inner valence 
(conduction) band regions in Fig. 5 (c).

\section{Zigzag nanotubes}

In this section, we discuss typical behaviors of disorder
effects in metallic zigzag nanotubes.  As an example, 
the (9,0) geometry is considered in the calculations.
We take the carbon number $N=3600$ here.  As the number
of carbons in the unit cell is 36, the system is composed
of 100 unit cells.

Figs. 6 (a) and (b) shows the entire DOS and the enlarged
DOS near the Fermi energy of the clean (9,0) nanotube.
Again, several one dimensional peaks are present in both
figures.  Figure 6 (c) shows the electronic conductance
in the units of $2e^2/h$ of the energy region $-1.6 t \leq
E \leq 1.6t$.  We observe several broad peaks of the conductance
at the energies $E = \pm 0.5t, \pm 0.6t, \pm 1.0t$, and $\pm 1.3t$, 
for example.  This is again typical results of the calculations
by the Thouless formula.  The conductance in the inner
valence (conduction) band regions is around 1.2 $(2e^2/h)$,
and that of the energy region around the Fermi energy
is about 0.9 $(2e^2/h)$.  The discrete peaks are due to
the discrete energy meshes, and they are of no importance.

Next, we discuss disorder effects on the DOS and electronic
conductance by taking results of one sample data of disorder
sets.  In Figs. 7 (a) and (b), the one dimensional sharp
peaks broaden as well by the disorder of the strength
$t_s = 0.15t$.  However, the magnitude of the DOS around
the Fermi energy $E=0$ does not change its value.
Figure 7 (c) shows the electronic conductance of the
same disorder sample.  The magnitude of the conductance
in the inner valence (conduction) band regions is around
0.3 $(2e^2/h)$, and this is apparently decreased from
that of the clean system.  But, the conductance near 
the Fermi energy is about 0.5 $(2e^2/h)$, and is of
the same order of the magnitudes with that of the 
clean system.   This is due to the relatively extended
nature of the wave functions around the Fermi energy.

Therefore, the qualitative properties of disorder effects
on the DOS and the electronic conductivity do not depend
on whether the carbon nanotubes are armchair or zigzag type.

\section{Disorder strength dependence}

In the previous sections, we have looked at the disorder
effects by showing the DOS and the conductance data
of one disorder sample with $t_s = 0.15t$ for the three
geometries of metallic carbon nanotubes.  In this section, 
we concentrate on the conductance at the Fermi energy
$E=0$, and look at the dependence on the disorder strength 
$t_s$.  Sample average is taken over 100 disorder samples.
We have taken relatively larger system sizes, so we
cannot take larger sample numbers.  However, the average
data seem relatively smooth in order to consider disorder
strength dependences.

Figure 8 shows the average conductance at $E=0$ as a function
of $t_s$.  The squares, circles, and triangles are for
(5,5), (10,10), and (9,0) nanotubes, respectively.
The conductances at $t_s=0$ are about 0.5, 0.6, and 0.9,
in units of $2e^2/h$ for (5,5), (10,10), and (9,0) nanotubes.
The magnitude at $t_s = 0.15t$ is at about 0.3 $(2e^2/h)$
for the three plots.  Therefore, the conductance of (5,5)
nanotube decreases by the factor about $1/1.6$.  The conductance
of the (10,10) nanotube decreases by the factor about $1/2$.
The conductance of the (9,0) nanotube decreases by the factor
about $1/3$.  Thus, the electronic conductance of the realistic
system with thermal fluctuations of phonons might be decreased
by the factor $1/2 \sim 1/3$, naturally.

The extended nature of electronic states at the Fermi energy
will contribute to several interesting transport properties
observed in experiments.  The ballistic conduction [14]
and the quantum single electron tunneling [12] are several
examples of the recent experiments of carbon nanotubes.  
We expect further developments of experimental transport
studies, which will promote theoretical investigations of 
carbon nanotubes as well.

\section{Summary}

Disorder effects on density of states and electronic
conduction in metallic carbon nanotubes have been analyzed
by a tight binding model with Gaussian bond disorder.
Metallic armchair and zigzag nanotubes have been considered.
We have obtained a conductance which becomes smaller by the 
factor $1/2 \sim 1/3$ from that of the clean 
nanotube.  This decrease mainly comes from lattice 
fluctuations of the width which is comparable to thermal
fluctuations.   We have also found that the suppression of 
electronic conductance around the Fermi energy due to
disorder is smaller than that of the inner valence
(and conduction) band states.  This is due to the extended
nature of electronic states around the Fermi energy

\mbox{}

\begin{flushleft}
{\bf Acknowledgements}
\end{flushleft}

\noindent
Useful discussion with the members of Condensed Matter
Theory Group\\
(\verb+http://www.etl.go.jp/+\~{}\verb+theory/+),
Electrotechnical Laboratory is acknowledged.  
The author specially thanks Barry Friedman for reading
the original manuscript critically.  Numerical 
calculations have been performed on the DEC AlphaServer 
of Research Information Processing System Center (RIPS), 
Agency of Industrial Science and Technology (AIST), Japan.

\pagebreak
\begin{flushleft}
{\bf References}
\end{flushleft}

\noindent
$[1]$ M. S. Dresselhaus, G. Dresselhaus, and P. C. Eklund,
``Science of Fullerenes and Carbon Nanotubes",
(Academic Press, San Diego, 1996).\\
$[2]$ R. Saito, G. Dresselhaus, and M. S. Dresselhaus,
``Physical Properties of Carbon Nanotubes",
(Imperial College Press. London, 1998).\\
$[3]$ J. W. Mintmire, B. I. Dunlap, and C. T. White,
Phys. Rev. Lett. {\bf 68}, 631 (1992).\\
$[4]$ N. Hamada, S. Sawada, and A. Oshiyama,
Phys. Rev. Lett. {\bf 68}, 1579 (1992).\\
$[5]$ R. Saito, M. Fujita, G. Dresselhaus, and M. S. Dresselhaus,
Appl. Phys. Lett. {\bf 60}, 2204 (1992).\\
$[6]$ K. Tanaka, K. Okahara, M. Okada, and T. Yamabe,
Chem. Phys. Lett. {\bf 193}, 101 (1992).\\
$[7]$ K. Harigaya, Phys. Rev. B {\bf 45}, 12071 (1992).\\
$[8]$ K. Harigaya and M. Fujita, Phys. Rev. B {\bf 47},
16563 (1993).\\
$[9]$ J. W. G. Wild\"{o}er, L. C. Venema, A. G. Rinzler,
R. E. Smalley, and C Dekker. Nature {\bf 391}, 59 (1998).\\
$[10]$ T. W. Odom, J. L. Huang, P. Kim, and C. M. Lieber,
Nature {\bf 391}, 62 (1998).\\
$[11]$ L. Langer et al, Phys. Rev. Lett. {\bf 76},
479 (1996).\\
$[12]$ M. Bockrath et al, Science {\bf 275}, 1922 (1997).\\
$[13]$ S. J. Tans et al, Nature {\bf 394}, 761 (1998).\\
$[14]$ S. Frank, P. Poncharal, Z. L. Wang, and W. A. de Heer,
Science {\bf 280}, 1744 (1998).\\
$[15]$ C. T. White and T. N. Todorov, Nature {\bf 393},
240 (1998).\\
$[16]$ R. Landauer, Philos. Mag. {\bf 21}, 863 (1970).\\
$[17]$ W. Tian and S. Datta, Phys. Rev. B {\bf 49}, 5097 (1994).\\
$[18]$ L. Chico, L. X. Benedict, S. G. Louie,
and M. L. Cohen, Phys. Rev. B {\bf 54}, 2600 (1996).\\
$[19]$ L. Chico, M. P. L\'{o}pez Sancho, and M. C. Mu\~{n}oz,
Phys. Rev. Lett. {\bf 81}, 1278 (1998).\\
$[20]$ J. T. Edwards and D. J. Thouless, J. Phys. C {\bf 5}, 
807 (1972).\\
$[21]$ D. C. Licciardello and D. J. Thouless, J. Phys. C 
{\bf 11}, 925 (1978).\\
$[22]$ K. Harigaya, Phys. Rev. B {\bf 48}, 2765 (1993).\\
$[23]$ M. S. Dresselhaus, G. Dresselhaus, and R. Saito,
Phys. Rev. B {\bf 45}, 6234 (1992).\\

\pagebreak

\begin{flushleft}
{\bf Figure Captions}
\end{flushleft}

\mbox{}

\noindent
Fig. 1.  Possible way of making helical and nonhelical tubules.
The open and closed circles indicate the metallic and semiconducting
behaviors of the tight binding model, respectively.  The Kekul\'{e}
structure is superposed on the honeycomb lattice pattern.

\mbox{}

\noindent
Fig. 2. Density of states (DOS) and electronic conductance of
the clean (5,5) nanotube.  Figures 2 (a) and (b) show the entire
DOS and the enlarged DOS of the energy region $-1.5 t \leq E
\leq 1.5t$, respectively.  Figure 2 (c) shows the electronic 
conductance in units of $2e^2/h$ of the low energy regions.

\mbox{}

\noindent
Fig. 3. Density of states (DOS) and electronic conductance of
one sample of the (5,5) nanotube with the disorder strength
$t_s = 0.15t$.  Figures 3 (a) and (b) show the entire
DOS and the enlarged DOS of the energy region $-1.5 t \leq E
\leq 1.5t$, respectively.  Figure 3 (c) shows the electronic 
conductance in units of $2e^2/h$ of the low energy regions.

\mbox{}

\noindent
Fig. 4. Density of states (DOS) and electronic conductance of
the clean (10,10) nanotube.  Figures 4 (a) and (b) show the entire
DOS and the enlarged DOS of the energy region $-1.5 t \leq E
\leq 1.5t$, respectively.  Figure 4 (c) shows the electronic 
conductance in units of $2e^2/h$ of the low energy regions.

\mbox{}

\noindent
Fig. 5. Density of states (DOS) and electronic conductance of
one sample of the (10,10) nanotube with the disorder strength
$t_s = 0.15t$.  Figures 5 (a) and (b) show the entire
DOS and the enlarged DOS of the energy region $-1.5 t \leq E
\leq 1.5t$, respectively.  Figure 5 (c) shows the electronic 
conductance in units of $2e^2/h$ of the low energy regions.

\mbox{}

\noindent
Fig. 6. Density of states (DOS) and electronic conductance of
the clean (9,0) nanotube.  Figures 6 (a) and (b) show the entire
DOS and the enlarged DOS of the energy region $-1.5 t \leq E
\leq 1.5t$, respectively.  Figure 6 (c) shows the electronic 
conductance in units of $2e^2/h$ of the low energy regions.

\mbox{}

\noindent
Fig. 7. Density of states (DOS) and electronic conductance of
one sample of the (9,0) nanotube with the disorder strength
$t_s = 0.15t$.  Figures 7 (a) and (b) show the entire
DOS and the enlarged DOS of the energy region $-1.5 t \leq E
\leq 1.5t$, respectively.  Figure 7 (c) shows the electronic 
conductance in units of $2e^2/h$ of the low energy regions.

\mbox{}

\noindent
Fig. 8.  The average conductance at the Fermi energy $E=0$ 
as a function of $t_s$.  The squares, circles, and triangles 
show the numerical data of (5,5), (10,10), and (9,0) nanotubes, 
respectively.

\end{document}